\documentclass[twocolumn,showpacs,showkeys,preprintnumbers]{revtex4}

\usepackage{graphicx}
\usepackage{subfigure}
\usepackage{lscape}
\usepackage{dcolumn}
\usepackage{bm}
\usepackage{color}
\usepackage{amsmath} 
\usepackage{amssymb} 

\begin{document}

\title{Explaining the characteristics of lateral shower age of cosmic ray extensive air showers}

\author{Rajat K. Dey}
\email{rkdey2007phy@rediffmail.com}
\author{Animesh Basak}
\email{ab.astrophysics@rediffmail.com}
\affiliation{Department of Physics, University of North Bengal, Siliguri, WB 734 013 India}

\begin{abstract}
A simple analytical argument is proposed for a possible explanation of the characteristics of the lateral shower age ($s$) of proton ($p$)/nuclei-initiated showers. The analytical argument states that lateral density distribution (LDD) of electrons of a $p$-initiated shower is due to superposition of several electromagnetic (EM) sub-showers developed at a very early stage in the atmosphere from the decay of neutral pions ($\pi^{0}$s). Thanks to the superposition property of the electron LDD in a $p$ shower, a plausible analytical parametrization has been worked out by giving well represented analytic function for the electron LDDs of $p$- and $\pi^{0}$-initiated showers. Based on cosmic ray extensive air shower simulations, we have validated how the various characteristics of $s$ can be understood in the context of the present analytical argument. The $s$ parameter of a $p$ shower and its correlations with the shower ages of electron- and $\pi^{0}$-initiated showers supports the idea that the result of superposition of several EM sub-showers initiated by $\pi^{0}$s with varied energies at a very early stage might produce the LDD of electrons of a $p$ shower. It is also noticed with the simulated data that the stated feature still persists concerning the notion of the local shower age parameter.

\end{abstract}

\maketitle

\section{Introduction}

The study of the average longitudinal development of extensive air showers (EASs) initiated by pure electromagnetic (EM) components such as electrons ($\rm e^{-}$ or $e^{+}$) or photons ($\gamma$s) first perceived the necessity of the notion of the {\lq age\rq} of a shower [1]. Such a concept was applied later to the lateral distribution of shower electrons around the EAS axis [2-3]. It was soon realized from the EM cascade theory that every shower has to be assigned by an age, and it remains true for showers generated by hadrons or nuclei as well.

Some current and recent works have revisited several conceptual issues in connection with the shower age particularly for protons/nuclei generated showers [4-8] and references therein. Basically the trace of the shape/form of the lateral distribution of shower electrons ({\it  i.e.} $\rm e^{-}+e^{+}$) is indicated by the shower age in the EM cascade theory for showers generated by $e^{-}$/$e^{+}$/$\gamma$s, and is fairly valid also for hadron/nuclei initiated showers [9]. 

The EAS parameters that are obtained directly from air shower experiments are densities and arrival times of shower particles. Out of different charged secondaries in a shower, about $\geq 90\%$ are electrons $e^{-}+e^{+}$. The important shower parameters such as the lateral shower age $s$, the electron/shower size ($\rm N_e$), and the shower core ($\rm{x_{0},y_{0}}$) are obtained by fitting the electron lateral density data (LDD) with the Nishimura-Kamata-Greisen (NKG) type lateral density function (LDF) [2,10]. It was found from both the experimental and simulation data that the NKG function or other type of modified LDF with a single $s$ is not adequate to describe the LDD of electrons initiated by proton ($p$)/nuclei/$\gamma$-ray primaries at all radial distances from the EAS core accurately [5,7]. In the present work, we have noticed the above behaviour in the LDDs of electrons for simulated showers initiated by $e^{-}$s and $\pi^{0}$s as well. This indicates that the lateral shower age varies with the radial distance, which is in character the shower age at a point, and was labeled as the local age parameter (LAP) by Capdevielle and Gawin [11]. To assign a single age parameter to each shower, some works took the minimum value of the LAP from its variation against the radial distance [5,8] while some other works took some sort of averaging of different LAPs falling between the first minima and the second maxima [5,7-8,11]. 

It is noteworthy to mention that the $s$ parameter estimated from the above two distinctive methods is adjudged an important parameter for studying the primary cosmic rays (PCRs). These parameters have been used to correlate with other EAS observables to identify the nature of the shower initiating PCR particles [7-8]. The parameter also provides a basic relationship between the slope of the LDF and the crucial CR mass sensitive parameter \lq{the depth of shower maximum\rq} ($X_{\rm{max}}$) in the ultra-high-energy (UHE) range. The parameter $s$ characterizes the development stage of an EAS albeit the parameter manifests different radial dependency. The magnitude of $s$ is intimately connected with the degree of steepness of LDD of shower electrons. In the present work however we will not at all re-examine any of the aforesaid characteristics of the $s$ parameter by analyzing the simulation/observed data. This work will rather discuss how the different observed properties associated with $s$ can be understood more precisely with the proposed simple analytical argument and application of EAS simulations.

The LDD of electrons of $p$/nuclei-initiated shower is a result of superposition of several EM ($e\textendash\gamma$) sub-showers generated from the decay of $\pi^{0}\rightarrow \gamma + \gamma$ at the early stages of EAS development. The LDD of electrons of each EM sub-shower can be  represented by the well-known NKG type LDF, $\rho_{i}(r)=N_{e_{i}}C(s_{i})X^{s_{i}-2}{(1+X)^{s_{i}-4.5}}$. Guided by the superposition property indicated in the present analytical argument, we express the resultant LDD of electrons of $p$/nuclei-initiated shower by $\rho{(r)}=\sum_{i}{\rho_{i}(r)}$ with $\rho(r)=N_{e}C(s)X^{s-2}{(1+X)^{s-4.5}}$. A full description of this analytical argument together with the superposition property is followed by the next section.     

This paper is organized as follows. In section 2 we present the simple analytical argument in the quest for possible physical explanation of the observed behavior of lateral shower age. In section 3 the characteristics of the air shower simulation by the CORSIKA code are described. Results concerning the concept of lateral shower age obtained from the implementation of the analytical approach using simulation data are presented and discussed in section 4. Finally we give some conclusions in section 5.         

\section{Elements of the simple analytical approach}

The characteristics of the observed LDD of electrons produced from CR  $p$/nuclei are imperative for the understanding of the PCRs, and also the EAS development in the atmosphere. It is well known in the expert community that the EM content of EAS is formed by a superposition of partial EM sub-cascades initiated by $\gamma$-rays resulting from decays of (mostly) $\pi^{0}$s. The dominant contribution comes from those $\pi^0$s that are produced at very early stages of EAS development. In the cascade theory, a pure EM cascade can be expressed independently by a pair of shower age parameters [3,12]. One of them is known as the longitudinal shower age $s_{\parallel}$, and it is the saddle point in the inverse Melline transformation of cascade diffusion equations [6] and references therein. The parameter manifests the development stage of an EM cascade in the atmosphere. The other one by which the LDD of $e$/$\gamma$s is described, termed as the lateral shower age $s$, which indicates the slope of the NKG type LDF of an EAS [3]. In [12], it was shown that $s_{\parallel}=s$ for pure EM showers. The work of Kamata and Nishimura in [12] also validated that both the longitudinal and lateral structures of $e$/$\gamma$s of hadron generated showers can be described by a single cascade with an admissible value to the slope parameter. Their study however failed to interpret the longitudinal EAS development by the so called slope parameter which acts as an indicator of shower age parameter.

Many EAS experiments found that the $s$ parameter differs from the $s_{\parallel}$ for EASs generated by $p$/nuclei. These two age parameters were found to follow an approximate relation $s_{\parallel}-s\geq \delta$, with $\delta\approx 0.2$ [13-14]. In [15], authors established a relation $s_{\parallel}\sim 1.3s$ using simulations. Generally, an EAS experiment equipped with fluorescence/Cherenkov detectors could estimate the $s_{\parallel}$. On the contrary, the $s$ parameter can be obtained by fitting the electron LDD data received from scintillation detectors placed in an EAS array with the NKG type LDF. It should be however mentioned that the introduction of the LAP (discussed in section 1) instead of $s$ for better understanding of the radial dependency of shower age follows a different concept than what complied by $s$ or $s_{\parallel}$ [5].    

Now, with the help of the simple analytic route in this paper, we are aiming for deeper physical insights on the concepts of the parameter $s$ for showers generated by p/nuclei. For the effectiveness of the adopted analytic argument, the shower data obtained from the simulations by running the CORSIKA code [16] have been exploited.  

According to the present analytical argument, a hadron initiated shower is assumed to be a result of superposition of number of partial electron-photon sub-cascades started mostly from the decay of $\pi^{o}$s at different developmental stages of the shower in the atmosphere. Additionally, the major contribution in the observed e/particle density comes from the $e\textendash \gamma$ sub-cascades resulting from decays of (mostly) $\pi^{o}$s generated at the early stages. The LDD  of electrons of a particular $e\textendash \gamma$ sub-cascade (say, the i$^{th}$ sub-cascade) is believed to be described by the NKG type LDF with a slope/lateral age parameter $s_{i}$. Again the overall LDD of electrons of an EAS generated by p/nuclei can be well represented by a resultant NKG type LDF having a different slope/lateral shower age in various analyses. Hence, the superposition property applies to electron-photon sub-cascades in a shower requiring the following equality 

\begin{equation}
\begin{aligned}
N_{e}C(s)X^{s-2}{(1+X)^{s-4.5}} 
\\= \sum_{i}{|N_{e_{i}}C(s_{i})X^{s_{i}-2}  {(1+X)^{s_{i}-4.5}|}},
\end{aligned}
\end{equation} 

where $N_{e}$, being the electron size (the sum $e^{+}+e^{-}$) of the resultant electron LDD and, $N_{e_{i}}$ represents the electron size for the LDD of the i$^{th}$ electron-photon sub-cascade with shower age $s_{i}$. Let $\acute{s}$ denotes the lateral age of an equivalent EM cascade of the hadron shower which has been generated from primary $e$/$\gamma$. The EM cascade also possesses the electron size $N_{e}$ that is assumed to be equal to the electron size of the hadronic shower. Dividing both sides of eq.(1) by  $N_{e}C(\acute{s})X^{\acute{s}-2}\\{(1+X)^{\acute{s}-4.5}}$, being the LDF for the LDD of electrons of an equivalent EM cascade, one may then get the following equation after taking logarithm from both sides, 
	
\begin{equation}
s=\acute{s} - \frac{ln[C(s)/C(\acute{s})]-ln~\sum_{i}\alpha_{i}C(s_{i})/C(\acute{s})h^{\delta_{i}}}{ln(h)},~~~~~~~
\end{equation}
	
where $\alpha_{i}=N_{e_{i}}/N_{e}$, $h=X(1+X)$ with $X=r/r_{m}$ and $\delta_{i}=s_{i}-\acute{s}$.
	
It is expected that all the three lateral shower ages appear in eq.(2) should be different from one another but the normalization factors $C(s)$, $C(\acute{s})$ and $C(s_{i})$ do differ very negligibly. These factors $C(s)$, $C(\acute{s})$ and $C(s_{i})$ have the same functional form as $\frac{1}{2\pi}\frac{\Gamma(4.5-s)}{\Gamma(s)\Gamma(4.5-2s)}$ in the NKG function. $s$ parameters in all these $C(s)$s are obtained from the fit on the simulated densities of electrons of $p$-, $e^{-}$- and $\pi^{0}$-initiated showers using the NKG type LDF. We have obtained the average values of these factors $C(s)$, $C(\acute{s})$ and $C(s_{i})$ along with their uncertainties from the analysis of simulated data as $0.359\pm{0.033}$, $0.337\pm{0.020}$ and $0.330\pm{0.006}$. Taking $C(s)\approx C(\acute{s})\approx C(s_{i})$, we can rewrite eq. (2) as,
	
\begin{equation}
s\approx\acute{s} + \frac{ln~\sum_{i}\alpha_{i}h^{\delta_{i}}}{ln(h)}~~~~~~~
\end{equation}
	
Now, we suppose that all the secondary electron-photon sub-cascades (say, $n$ number of sub-cascades in total) have the same size $n_{e}$ (or more explicitly, $N_{{e}_{i}}\approx n_{e}$ for $i=1,2,3....n$), and also have the same age  ($s_{i}\approx {\tilde{s}}$, again for $i=1,2,3....n$). We assume that parameter $\delta_{i}$ accounts the difference between the lateral shower ages of two EM cascades, in which one refers to the i$^{th}$ electron-photon sub-cascade, and the rest is the effective EM cascade generated by primary $e$/$\gamma$. It is further assumed that $\delta_{i}$ will not change appreciably with the atmospheric depth, and the ratio ${N_{{e}_{i}}}/N_{e}$ will similarly not vary from one electron-photon sub-cascade to the other, then

\begin{equation}
s\approx\acute{s} + \frac{ln~[({n{n_{e}}}/N_{e})h^{\delta}]}{ln(h)}~~~~~~~
\end{equation}

where $\delta = \tilde{s}-\acute{s}$ and $nn_{e}\approx N_{e}$. Under the above circumstances we have directly from eq. (2) with a positive $\delta$, 

\begin{equation}
s\approx\acute{s} - \delta = 2\acute{s}-\tilde{s} ~~~~~
\end{equation}

The variation with the atmospheric depth, can be 

\begin{equation}
\frac{ds}{dt} =2\frac{d\acute{s}}{dt}-\frac{d\tilde{s}}{dt}>\frac{d\acute{s}}{dt} ~~~~~~~
\end{equation}

One can rewrite eq. (2) in an alternative way, in terms of lateral density of shower electrons:

\begin{equation}
ln{[\frac{\rho_{Had}(r)}{\rho_{EM}(r)}]} = ln(n\alpha_{e}\beta_{e})+(\tilde{s}-\acute{s})ln[h],~~~~~~~
\end{equation}

where $\alpha_{e}={\frac{N_{{e}_{i}}}{N_{e}}},......\approx \frac{n_{e}}{N_{e}}$ and $\beta_{e}= C(\tilde{{s}_{i}})/C(\acute{s}),.....\approx{C(\tilde{s})/C(\acute{s})}\approx 1$ with (i=1,2,....n), and $n$ represents the total number of electron-photon sub-cascades produced from a hadronic primary via $\pi^{0}$-decays. We have then,

\begin{equation}
(\acute{s}-\tilde{s})\approx \frac{ln(n\alpha_{e}) - ln{[\frac{\rho_{Had}(r)}{\rho_{EM}(r)}]}}{ln[h]}\approx \delta, ~~~~~
\end{equation}

$\delta$ can be estimated from the above relation by using simulations.

It has been reiterated that $s$ parameter is evaluated from the fit on the simulated densities of electrons with the NKG type LDF for any shower ($p$/$e^{-}$/$\pi^{0}$-initiated shower) over a certain radial distance range from the EAS core. However, it is noticed that the NKG function with a single $s$ cannot describe the observed/simulated LDD of EAS electrons accurately at all radial distances over the concerned radial range, which implies that the lateral shower age is a function of radial distance [5]. Later some modifications into the NKG LDF were incorporated but the radial dependency on $s$ could not be removed completely. From that viewpoint, the idea of LAP was emerged which is in character the point-wise value of $s$ [5,11]. The well-known LAP for hadron initiated showers can be expressed by the following [11]:

\begin{equation}
s_{local}^{Had}(i,j)=\frac{ln(F_{ij}X_{ij}^{2}Y_{ij}^{4.5})}{ln(X_{ij}Y_{ij})} ~~~~~~~
\end{equation}

For the equivalent EM cascade, the corresponding LAP is,

\begin{equation}
s_{local}^{EM}(i,j)=\frac{ln(\acute{F}_{ij}\acute{X}_{ij}^{2}\acute{Y}_{ij}^{4.5})}{ln(\acute{X}_{ij}\acute{Y}_{ij})} ~~~~~~~
\end{equation}

The present argument treats a hadronic shower as a superposition of n number of electron-photon sub-cascades, the LAP of the hadronic shower would therefore take the following form.

\begin{equation}
s_{local}^{Had}(i,j)\approx {\frac{ln(\sum_{k}\tilde{F}_{ij,k}\tilde{X}_{ij,k}^{2}\tilde{Y}_{ij,k}^{4.5})}{ln(\tilde{X}_{ij,k}\tilde{Y}_{ij,k})}}~~~~~~~
\end{equation}

It is now clear that, $\tilde{X}_{ij,1}=\tilde{X}_{ij,2},....=\acute{X}_{ij}=X_{ij}$ and $\tilde{Y}_{ij,1}=\tilde{Y}_{ij,2},....=\acute{Y}_{ij}=Y_{ij}$. Hence,

\begin{equation}
s_{local}^{Had}(i,j)\approx {\frac{ln((\sum_{k}\tilde{\rho}_{ij,k})\tilde{X}_{ij}^{2}\tilde{Y}_{ij}^{4.5})}{ln(\tilde{X}_{ij}\tilde{Y}_{ij})}}~~~~~~~
\end{equation}

where $\tilde{\rho}_{ij,k}$ and $\acute{\rho}_{ij}$ are the ratios of electron densities between two adjacent points. Here, $\tilde{\rho}_{ij,k}$ is associated with the k$^{th}$ sub-cascade, generated from $\pi^{o}$-decay in an hadronic shower, while $\acute{\rho}_{ij}$ accounts the effective EM-shower.\\

One may verify the relation $s-\acute{s}\approx \delta$ (eq. (5)) or $\acute{s}-\tilde{s}\approx \delta$ (eq. (8)) in the context of the LAP if we can express $s_{local}^{Had}(min)-\acute{s}_{local}^{EM}(min)\approx \delta$ and also the $\acute{s}_{local}^{EM}(min)-\tilde{s}_{local}(min)\approx \delta$. The minimum value of the LAP that is assigned to a shower, is discussed below.

Let us recall the feature of the LAP from both the simulations and observations in recent works [5,7-8]. It was found that the parameter or the LDD of shower electrons exhibits some kind of scaling behavior if one studies the variation of the parameter with the radial distance from the EAS core [5]. Instead of taking multiple local ages belonging to a shower, a systematically chosen single age parameter was found appropriate to deal with some important issues such as composition, energy {\it etc.} on the PCRs [5,7-8]. The minimum value of the LAP that was noticed at a particular radial distance between the first and the second maxima in the $s_{local}$ versus $r$ curves appears very effective in several recent works [7-8].    

\section{Characteristics of the air shower simulation and data analysis}

The simulated showers have been generated with the combination of interaction models; EPOS(LHC)-UrQMD-EGS4 [17-19] embedded in the detailed air shower code CORSIKA ver. 7.7401 [16] for the present analysis. The hadronic interaction model EPOS-LHC looks for interaction processes above the energy $80$~GeV/n, and below this energy the model UrQMD works for the hadronic processes. The EGS4 program package addresses the EM interactions in the cascade development. The CORSIKA showers have been generated at the KASCADE [10] level. The kinetic energy cut-offs for secondaries like hadrons, muons and electrons are set at 0.5, 0.025 and 0.005 GeV. The US-standard atmospheric model [20-21] is adopted for the generation of simulated showers. The shower events have been generated primarily for $p$ and $e^{-}$ primaries at a fixed primary energy $2\times 10^{6}$~GeV and at the zenith angle $0^o$. Initially MC showers have been generated for $p$ and $e^{-}$ primaries with about 1000 events for each primary. Required number of $\pi^{0}$ showers with varied energies have also been generated by maintaining some restrictions on different free parameters in the CORSIKA steering file. A more detail about the $\pi^{0}$ shower events is discussed in the following paragraph.\\
\begin{table*}
  \renewcommand*{\arraystretch}{2}
	\begin{center}
		\begin{tabular}
			{|c|c|c|c|} \hline
			$E_{\pi^{0}}$-range~[GeV]   & $N_{\pi^{0}}^{\rm Event-No}$    & $N_{f}$    & Simulated~$\pi^{0}$~events  \\ \hline
			
			$(0.01\textendash 0.99)\times 10^5$& $104.53$ & $1$    & $105$    \\ \hline
			
			$(1\textendash 1.99)\times 10^5$   & $4.01$   & $1$    & $4$      \\ \hline
			
			$(2\textendash 2.99)\times 10^5$   & $1.36$   & $11$   & $15$     \\ \hline
			
			$(3\textendash 3.99)\times 10^5$   & $0.80$   & $10$   & $8$      \\ \hline
			
			$(4\textendash 4.99)\times 10^5$   & $0.48$   & $25$   & $12$     \\ \hline
			
			$(5\textendash 5.99)\times 10^5$   & $0.16$   & $50$   & $8$      \\ \hline
			
			$(6\textendash 6.99)\times 10^5$   & $0.16$   & $50$   & $8$      \\ \hline
			
			$(7\textendash 7.99)\times 10^5$   & $0.48$   & $25$   & $12$     \\ \hline
		\end{tabular}
		\caption {Energies and event numbers for $\pi^{0}$ showers are set from charged pions information about a depth $\approx 75$~gcm$^{-2}$ from the top in an average $p$-initiated shower with energy 2 PeV.} 
	\end{center}
\end{table*}

In the simulation, the starting altitude of the primary particle $p$ and $e^{-}$ is set corresponding to the value of the keyword {\lq FIXCHI\rq} here. For these primaries, FIXCHI is set to $0$~gcm$^{-2}$ which refers the top of the atmosphere but at the immediate vicinity above the starting point of the EAS [16]. In {\it this analytical argument}, a hadronic shower is believed to be a superposition of a certain number of electron-photon sub-cascades. These sub-cascades might have produced mostly from some parent $\pi^{0}\rightarrow {\gamma} +{\gamma}$ decays just below the first interaction point of the shower. We have extracted the average first interaction length from the generated $p$ showers, and it takes a value $\approx 65$~gcm$^{-2}$ in the present analysis. Hence, the starting altitude (or FIXCHI) of the parent particles $\pi^{0}$ that induce electron-photon sub-cascades is set judiciously a little below the first interaction point. From the analysis, we have found that FIXCHI$\approx 75$~gcm$^{-2}$ is appropriate to deliver better results.\\ 

In the primary $p$-air collisions, copious mesons are produced and majority of them are pions. The superposition property of the present analytical argument requires information on the total number of produced $\pi^{0}$ and their energy distribution from the analysis of $p$-initiated showers at the level around $75$~gcm$^{-2}$ from the top of the atmosphere. But a $\pi^{0}$ can propagate only a very short distance, about $\sim 25$~nm in air due to their short life time $\sim 10^{-16}$~sec. So these $\pi^{0}$ particles can't be counted at any given height in the simulation. They are rather tracked only by the analogy with the charged pions; all the species of pions ($\pi^{0}$, $\pi^{\pm}$) are believed to be generated with equal numbers and $\pi^{+}$ and $\pi^{-}$ numbers are known from the simulation which then fixes the required $\pi^{0}$ number. Knowing the 3-momenta of charged pions, the energy distribution of $\pi^{0}$ can be found from the stated analogy. Frequency distribution of energies of $\pi^{+}$ or $\pi^{-}$ will fix the $\pi^{0}$ numbers with different energies. Finally, the required number of $\pi^{0}$ showers are simulated corresponding to different energy ranges by setting $\rm{FIXCHI}=75$~gcm$^{-2}$ in CORSIKA steering file. We have then combined the LDDs of electrons contributed by all the $\pi^{0}$ showers at the ground level for describing the LDD of an average p shower eventually.\\

It has been reiterated that the dominant contribution to the EM content of an EAS comes from the very early stage of its development. In $p-\rm{air}$ collisions, most of the secondary pions ($\pi^{0}$, $\pi^{+}$, $\pi^{-}$) are produced in the central rapidity region, and are therefore characterized by small fractions of energy, taken from their parent hadrons. The largest contributions to the EM shower content come from {\lq hard\rq} $\pi^{0}$s produced, i.e., those which acquire $\geq 10$\% of the energy of the parent hadron. It is clear that the acquired energies ($E_{\pi^{0}}$) by the $\pi^{0}$s are unequal. Hence, we set the event number for $\pi^{0}$-initiated sub-showers from the frequency distribution of average charged pions (i.e., $N_{\pi^{0}}^{\rm event-no}=\frac{(\pi^{+}+\pi^{-})}{2}$) corresponding to different energy ranges. In Table 1, the unequal energies acquired by the secondary charged pions of an average p-initiated shower (energy; $\rm E_{p}=2$~PeV) are shown. The 2nd column of Table 1 accounts the $N_{\pi^{0}}^{\rm event-no}$ derived from the generated $\pi^{\pm}$ numbers around $75$~gcm$^{-2}$ from the top of the atmosphere from an average p-initiated shower. To simulate $\pi^{0}$ sub-showers, we convert the fractions in column 2 of Table 1 into integers by multiplying different numbers ($N_{f}$). The lateral densities so obtained from these events (column 4) at the ground level are divided finally by respective $N_{f}$s.

\section{Results concerning the properties associated with the lateral shower age}

Densities of EAS electrons are measured at different radial distances from the shower core. In the work, we first discuss how the variation of the simulated electron densities with the radial distance for $p$ generated showers can be understood with the proposed simple arguments. Next, we briefly discuss how the correlation between $s$ and $s_{\parallel}$ of the $p$ generated showers can be understood with the present approach. We will then take the main issue of the work how the properties associated with the slope (also being an indicator of lateral shower age [6]) of the LDF describing the simulated LDD of electrons for $p$ generated showers can be understood with this analytical argument. The method gets a strong basis while this analytical argument has been fruitfully applied to the radial behavior of the LAP of EASs.   

\subsection{Lateral density distribution of electrons in simulated showers}

It has been anticipated that the present argument might be effective for explaining the behavior of the parameter $s$, if the method has a direct impact on the LDDs of electrons for primary $p$, $e^{-}$ and $\pi^{0}$ showers. Moreover, the LDD data of EAS electrons are the basic information from which the shower data analysis starts for obtaining other important EAS parameters.

\begin{figure}
	\centering
	\includegraphics[width=0.5\textwidth,clip]{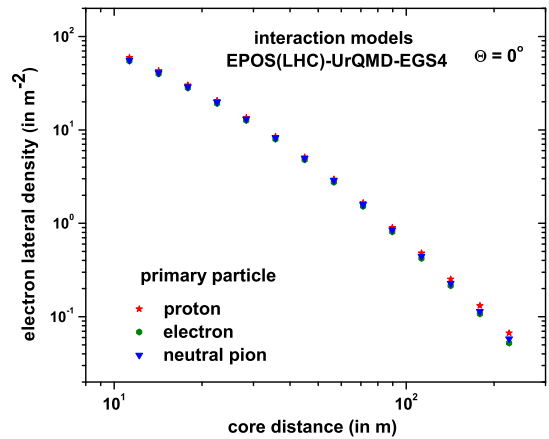} \hfill 
	\caption{Mean lateral density distribution of electrons for showers generated by $p$, an equivalent average EM cascade initiated by $e^{-}$, and the required number of $\pi^{0}$ sub-cascades.}
\end{figure} 

The simulated densities for secondary electrons at different radial distances from the shower core ($10 - 225$~m) are estimated in the analysis for $p$, an equivalent average EM cascade initiated by $e^{-}$ and $\pi^{0}$ showers at the KASCADE level. In the Fig. 1, we have plotted the mean electron densities over the above radial range for the three types of simulated showers. For the stated three types of showers, a very good agreement among the LDDs of shower electrons is noted over the radial distance range $10 - 100$~m. It ensures the fact that the average LDD of electrons of $p$-initiated showers can be well represented by the superposition of a justifiable number of electron-photon sub-cascades from the decay of secondary $\pi^{0}$s produced after the first interaction point of the same $p$ shower. At distances beyond $100~$m a very little contrast in the LDDs of electrons develops in the $p$ showers with respect to the combined $\pi^{0}$ sub-showers and the equivalent EM ($e^{-}$) shower. This might be due to the transverse momentum which the charged pions/kaons get in the first interaction. This actually gives a hint that the slope parameters of these LDDs of EAS electrons initiated by an average $p$/$e^{-}$, and required $\pi^{0}$s, might have a little difference from each other. The generic features of showers (e.g. shower-to-shower fluctuations) generated by different types of primary particles might be the other reason for that non-vanishing difference among the slope parameters of their LDDs even though these showers were assumed to be equivalent to each other. This work reviews though not show up in the paper the effectiveness of the NKG type function in describing the simulated LDD data of electrons generated from $p$, $e^{-}$ and $\pi^{0}$s. It is found that the NKG function uni-vocally describes the simulated LDD data of electrons irrespective of all these above primaries.       

\subsection{Correlation between the lateral and longitudinal shower ages}

The simulated LDD data of electrons at different radial distances are fitted usually by the NKG function for obtaining the parameter $s$. In [9,22], the longitudinal shape/age parameter $s_{\parallel}$ of the energy and angle distributions of shower electrons of very high-energy to UHE MC showers is well approximated to a simple form as

\begin{equation}
s_{\parallel} = \frac{3 X} { X + 2 X_{\rm max}},
\end{equation}

where $X_{\rm max}$ is the depth of shower maximum and $X$ is the atmospheric slant depth of the KASCADE level (in unit of radiation lengths). Knowing both the depth parameters the $s_{\parallel}$  of a shower can be estimated. Hence, one should use $s_{\parallel}$ instead of the depth $\rm X$ referring to the shower maximum. The value of $\rm X/X_{max}$ is the key for the description of the EAS development stage. This work however studies the difference between $s_{\parallel}$ and $s$ for $p$, $e^{-}$ and required number of $\pi^{0}$ generated simulated  showers. For these showers, the frequency distributions of differences between $s_{\parallel}$ and $s$ are shown in the fig. 2(a), (b) and (c). The present MC simulation gives the frequency distribution peaks within the range $\approx 0.50 - 0.55$, which is consistent with some early MC simulation predictions satisfying a general relation of the form, $s_{\parallel} - s\approx 0.5$  or equivalently $s_{\parallel} \sim 1.4{s}$ [15]. We also notice that the fluctuations ($\sigma_{s}$) of the difference between $s_{\parallel}$ and $s$ are sensitive to the nature of the shower initiating primary species. The correlations between $s_{\parallel}$ and $s$ reveal that an average $p$ shower at least can be explained as the result of superposition of several EM sub-showers generated by $\pi^{0}s$.    
\begin{figure}
	\centering
	\includegraphics[width=0.45\textwidth,clip]{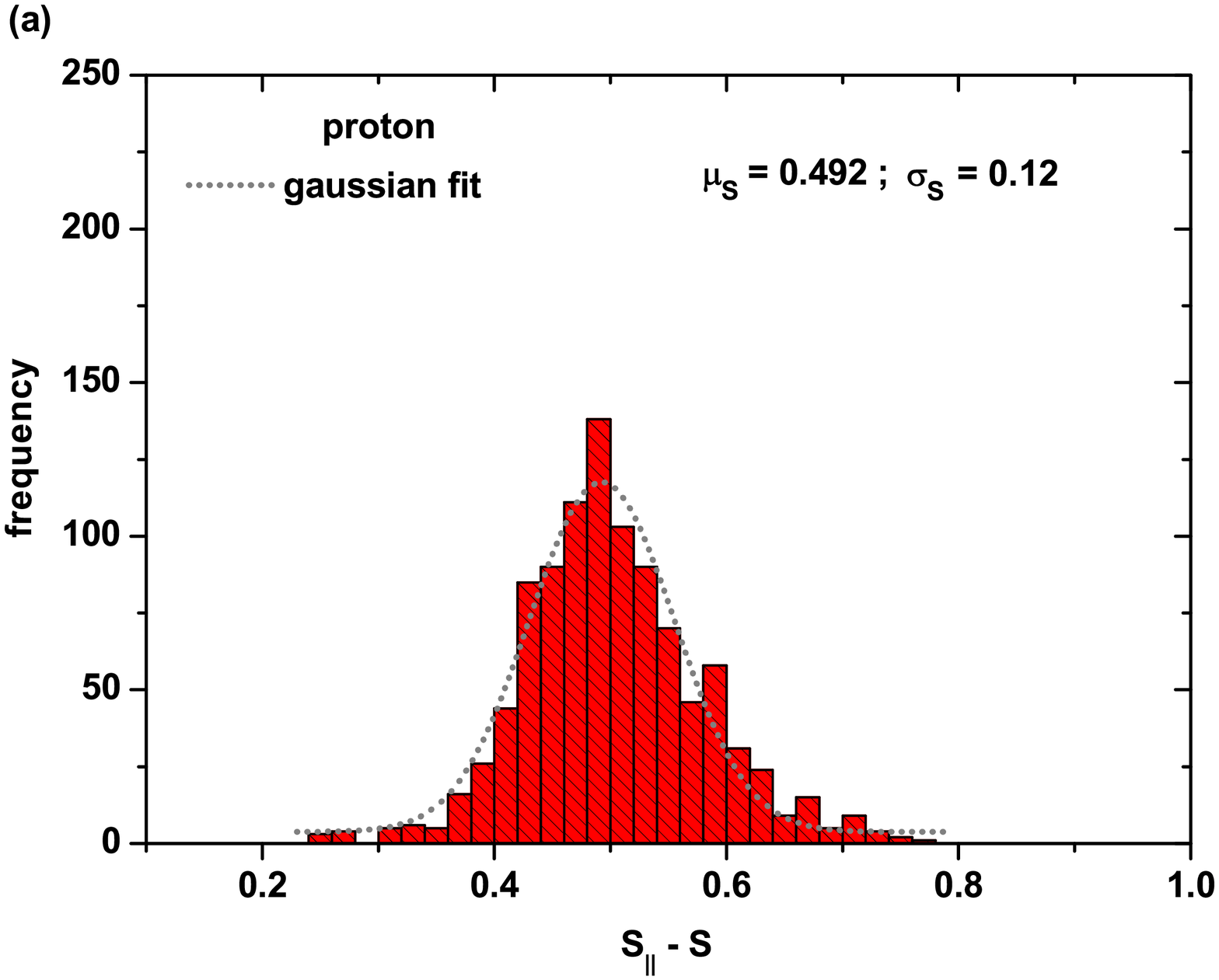} \hfill 
	\includegraphics[width=0.45\textwidth,clip]{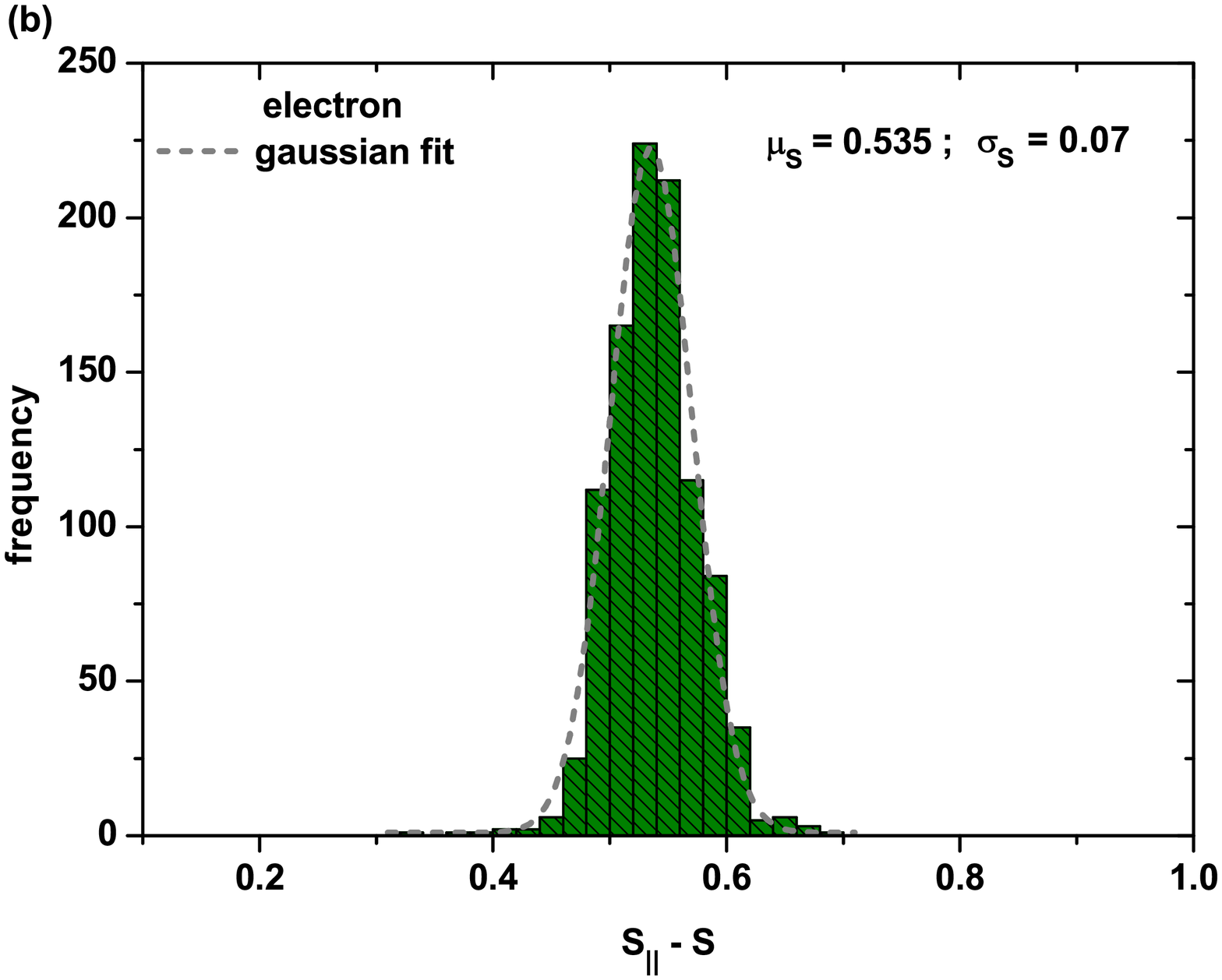} \hfill
	\includegraphics[width=0.45\textwidth,clip]{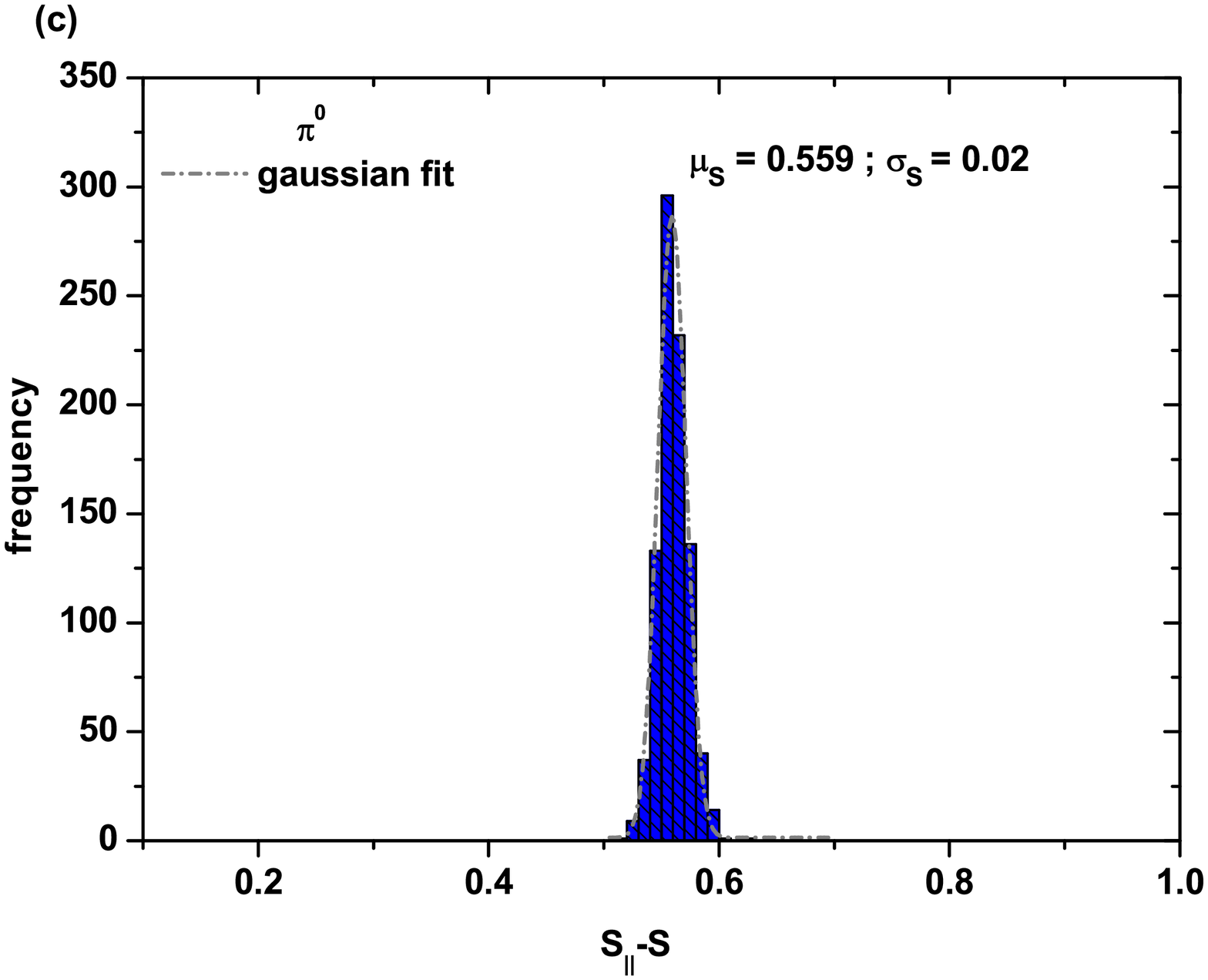} \hfill
	\caption{Distributions of differences between $s_{\parallel}$ and $s$ at the KASCADE site  for simulated (a) $p$, (b) $e^{-}$ and (c) $\pi^{0}$  showers.}
\end{figure}

\begin{table*}
	  \renewcommand*{\arraystretch}{2}
	\begin{center}
		\begin{tabular}
			{|c|c|c|c|c|c|c|} \hline
			$dt$~in~gcm$^{-2}$   & $ds$    & $d\acute{s}$    & $d\tilde{s}$    & $\frac{ds}{dt}$    & $\frac{d\acute{s}}{dt}$    & $\frac{d\tilde{s}}{dt}$    \\ \hline 
			
			$50$   & $0.0100$   & $0.0109$   & $0.0080$   & $2\times{10^{-4}}$   & $2.18\times{10^{-4}}$   & $1.60\times{10^{-4}}$  \\ \hline
			
			$50$   & $0.0144$   & $0.0102$  & $0.0172$    & $2.88\times{10^{-4}}$   & $2.04\times{10^{-4}}$   & $3.44\times{10^{-4}}$ \\ \hline
			
			$72$   & $0.0091$   & $0.0197$  & $0.0156$    & $1.26\times{10^{-4}}$   & $2.73\times{10^{-4}}$   & $2.17\times{10^{-4}}$  \\ \hline
		\end{tabular}
		\caption {Variation of lateral shower age with atmospheric depth. Row 2 , 3 and 4 are respectively for the depth variation intervals, $850 - 900$, $900 - 950$ and $950 - 1022$~gcm$^{-2}$.} 
	\end{center}
\end{table*}

\begin{figure}
	\centering
	\includegraphics[width=0.5\textwidth,clip]{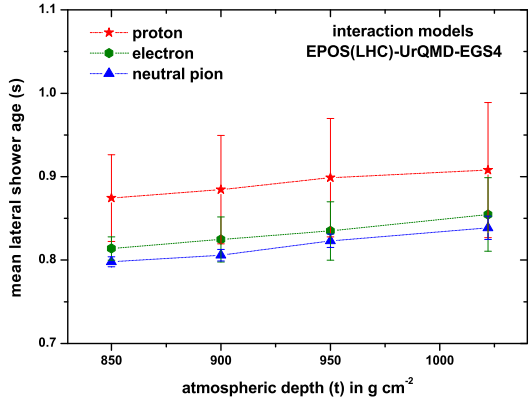} \hfill 
	\caption{Variation of mean $s$, $\acute{s}$ and $\tilde{s}$ with atmospheric depth at KASCADE level. Lines are only guiding our eyes.}
\end{figure} 

\subsection{Analytical description of the lateral shower age}

Having recognized this simple analytical argument for explaining the  behaviour of $s$ through eq. (2) - (8), we will now examine how well the LDD of electrons of simulated $p$-, $e^{-}$- and $\pi^{0}$-initiated showers follow them. In section 2, we have described how one could find the eq. (5) starting from eq. (3). From the analysis of the simulated LDD data of $p$-, $e^{-}$- and the required number of $\pi^{0}$-initiated  showers, we have obtained the parameter $\delta\approx s-\acute{s}=0.053$ and $2\acute{s}-\tilde{s}=0.87\approx s$. Shower ages are estimated by applying the fitting procedure (by virtue of a chi-square minimization routine using gradient search technique) to the simulated LDD data of electrons with the NKG LDF irrespective of primary particles. We have also checked the variation of $s$ with the atmospheric depth using simulated LDD data following the eq. (6). For the purpose, we have generated simulated showers at three different atmospheric levels ($850$, $900$, $950$~gcm$^{-2}$), in addition to the usual ground level $1022$~gcm$^{-2}$ of the KASCADE site. Results obtained from the shower data analysis on the variation of mean lateral shower age with the atmospheric depth is shown in fig. 3. The figure reveals that the shower age increases with atmospheric depth irrespective of $p$-, $e^{-}$- and $\pi^{0}$-initiated showers, which is a generic feature of shower development. 

From Table 2, we obtain $(2.\frac{d\acute{s}}{dt}-\frac{d\tilde{s}}{dt})=2.76\times{10^{-4}}$~cm$^{2}\rm{g^{-1}}$ for the atmospheric depth variation between 850 and 900~gcm$^{-2}$, and for the atmospheric depth interval $900-950$~gcm$^{-2}$, the value is nearly $6.4\times{10^{-5}}$. Finally for the atmospheric depth interval $950-1022$~gcm$^{-2}$, we have obtained the value $3.29\times{10^{-4}}$. For the depth variations $850-900$ and $950-1022$~gcm$^{-2}$, the condition $(2.\frac{d\acute{s}}{dt}-\frac{d\tilde{s}}{dt})> \frac{d\acute{s}}{dt}$ predicted in the adopted analytical method, is obeyed by the simulated showers.

The work illustrated so far involves the fit procedure for the estimation of the $s$ parameter for calculating $s_{\parallel} - s$, $\delta$, $\frac{ds}{dt}$ etc. An alternative attempt has been made for estimating the $\acute{s}-\tilde{s}\approx{\delta}$ directly via the eq. (8). The average densities of shower electrons for EM- and $p$-initiated showers at a radial distance around $50$~m from the EAS core are $\rho_{\rm EM}\approx 3.7368$ and $\rho_{\rm p}\approx 3.9342$~m$^{-2}$. The corresponding electron sizes are $N_{e}(EM)\approx N_{e}(\rm{p})\approx 1.9726\times{10^5}$ and $N_{e}(\pi_{0})\approx nn_{e}(\pi^{0})\approx 2.0337\times 10^5$. The value of $\frac{1}{ln[h]}$ is $\approx -2.4168$ with the Moli$\acute{e}$re radius $r_{m}=110$~m. Using all these simulated results in eq. (8), the value of $\delta\approx \acute{s}-\tilde{s}$ takes the value $\approx 0.05$. It deviates from the previous one i.e. $\delta\approx s-\acute{s}\approx 0.053$ typically by $\sim 5.5\%$ with the simulated data.   

\subsection{Compatibility of the present analytical argument with the local age parameter}    
This analytical argument may receive more impetus if we compute the parameter $\delta$ in terms of the LAPs introduced in section 2 for simulated $p$-, $e^{-}$- and $\pi^{0}$-initiated showers. In the analysis, the LAP for the LDD data was computed for each shower by applying eq. (9), (10) and (11) or (12) respectively for $p$-, $e^{-}$- and the required  number of $\pi^{0}$-initiated showers. It should be mentioned that the local age obtained for the LDD of an average $\pi^{0}$-initiated shower in a particular radial bin, say ($\rm r_{i},r_{j}$) would coincide with the LAP computed from the required number of $\pi^{0}$-initiated showers. For computing the local age from the required number of $\pi^{0}$-initiated sub-showers we will just take the ratio of the sums of electron densities ($\sum_{k}\rho_{ij,k}=\sum_{k}\frac{\rho_{i,k}}{\rho_{j,k}}$) of different $\pi^{0}$ showers between the points, :$r_{i},r_{j}$ in eq. (12). On the other hand, the LAP computed from the electron LDD data of an average $\pi^{0}$-initiated shower involves the ratio $\sum_{k}\rho_{ij,k}=\sum_{k}(\frac{\rho_{i,k}/n}{\rho_{j,k}/n}$) for $n$ number of equivalent $\pi^{0}$ showers.

\begin{figure}
	\centering
	\includegraphics[width=0.5\textwidth,clip]{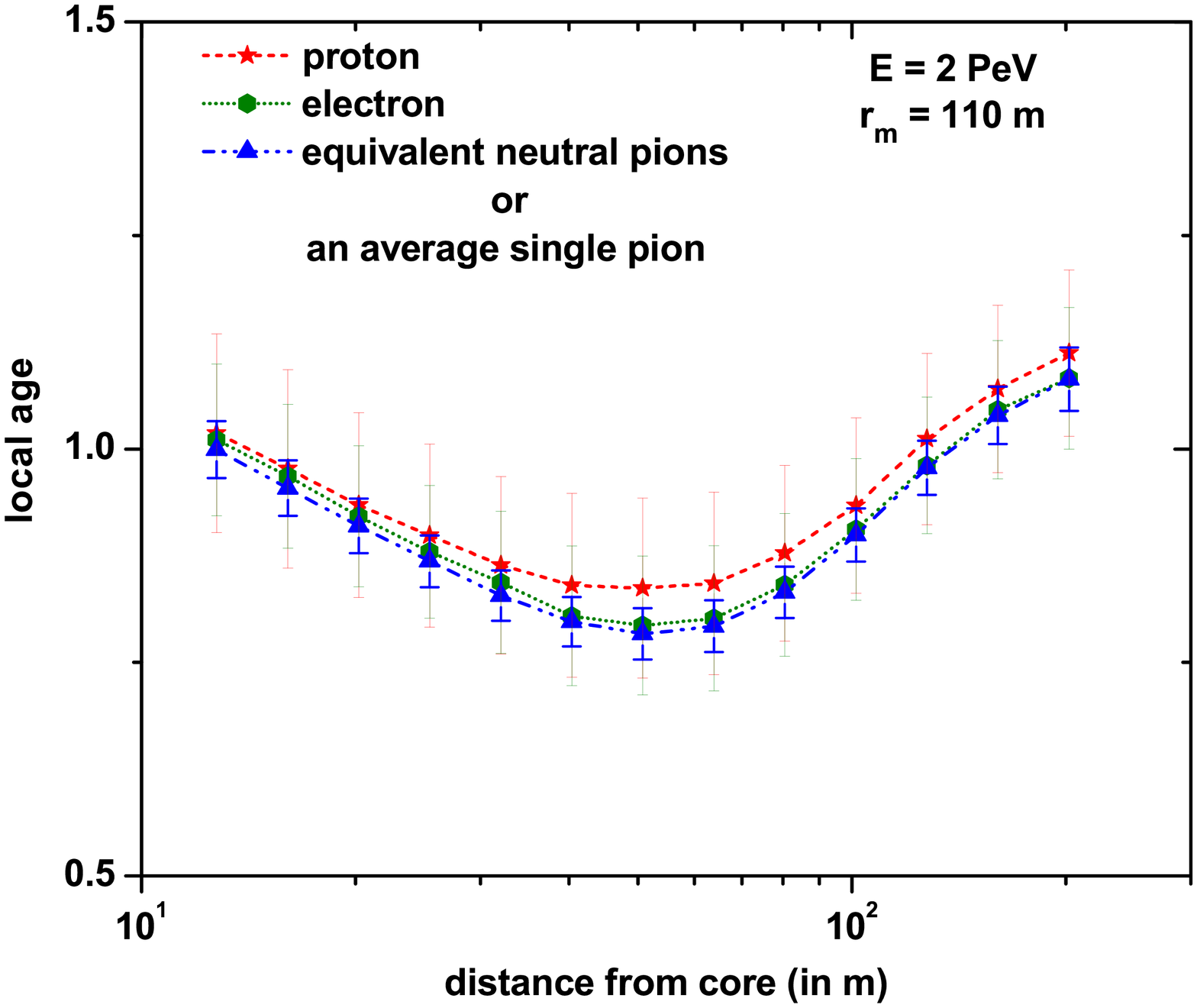} \hfill 
	\caption{Variation of the LAP computed from the simulated electron LDD data with radial distance for $p$, $e^{-}$, required number of $\pi^{0}$ showers or an average $\pi^{0}$ shower. Lines are only guiding our eyes.}
\end{figure}

The variation of the LAP with the radial distance is shown in Fig. 4. The error of the LAP arises due to density fluctuations of simulated electron data and the uncertainties in radial distance estimation remains within $0.04$ for $12<r<205~$m. Here, the minimum value of the LAP in the present analysis is the LAP value at the radial distance about $50~$m with the $r_{m}=110~$m, and is assigned to a shower as the lateral shower age parameter [5,8]. With the simulated data, we have checked the parameter $\delta$ using this alternatively defined lateral shower age parameter. We have obtained $\delta\approx s_{local}(min)-\acute{s}_{local}(min)\approx 0.044$ and $\delta \approx \acute{s}_{local}(min)-\tilde{s}_{local}(min)\approx 0.01$. In this local age representation, these two $\delta$ values differ much from each other. We have also noticed that $\delta$ increases with the radial distance from the EAS core. 

\section{Conclusions}
Investigating in detail the characteristics of the lateral shower age with simulated shower data, we conclude the following from the present analysis. 

The simulated electron LDDs of $p$-, $e^{-}$- and $\pi^{0}$-initiated showers unequivocally support the idea, explained in the adopted  analytical argument.

The candidature of $s$ as a crucial EAS parameter of the longitudinal development of a shower is noticed.

The difference in the numerical values between the lateral and longitudinal shower ages can be explained by the superposition property obeyed  by a number of EM sub-showers initiated by $\pi^{0}$s mostly from a predetermined atmospheric level at the KASCADE site. Moreover, the frequency distributions of these differences support the fact that the longitudinal and lateral developments of a shower are closely correlated by their shape parameters. The shape of these distributions and their spreads are found sensitive to CR mass composition.

The numerical value of the difference $\delta\approx s-\acute{s}=0.053\neq 0$ indicates that the radial dependence of $s$ is different than that of $\acute{s}$, because of the 2nd term in eq. (2). This otherwise supports the important underlying fact (superposition property) of this work which indeed could explain the behavior of the shape parameter of a $p$/nuclei initiated shower. The result $\acute{s}<s$ reveals the fact that for pure EM cascade, the LDD is steeper than that of the hadronic cascade. Applying simulated data to the eq. (8), we have obtained $\delta\approx \acute{s}-\tilde{s}=0.05$, is in agreement with the fact that $\pi^{0}$-initiated sub-shower is steeper than that of a pure EM shower. For the variation of the atmospheric depth, the condition $(2.\frac{d\acute{s}}{dt}-\frac{d\tilde{s}}{dt})>\frac{d\acute{s}}{dt}$ is found to valid in two atmospheric depth intervals; $800 - 850$ and $900-1022~$gcm$^{-2}$ by simulated data.

The radial variation of the LAP has reiterated the fact that a single value of $s$ is inadequate to describe the LDD of electrons accurately by a LDF (NKG) in the entire radial distance from the EAS core. The nature of variation of the LAP for $p$-, $e^{-}$- and $\pi^{0}$-initiated showers are portrayed as a generic feature of the LDDs of electrons in showers. A more rational single-valued minimum LAP is assigned to each shower for the purpose of using it as a good estimator in CR studies. The value of $\delta$ is almost recovered in the language of LAP (i.e. $s_{local}(min)-\acute{s}_{local}(min)$) but the value $\acute{s}_{local}(min)-\tilde{s}_{local}(min)$ deviates much from its earlier value in terms of lateral shower age obtained from fitting procedure.

It is noteworthy to mention that the paper has put one of the main aspects of the work to extreme: considering a fixed number of $\pi^{0}$s with varied energies produced at the same atmospheric depth (Table 1). It is more justifiable from the pragmatic point of view that one has to account for at least $2-3$ steps for the respective hadronic cascades. We have checked this aspect, and observed that the produced $\pi^{0}$s at the lower atmospheric depths do not have noticeable contribution to reproduce the LDD for $p$-induced EAS because these $\pi^{0}$s could pick up a very little energy from high-energy hadronic interactions.    

\section*{Acknowledgment}
Authors sincerely thank Dr. T. Pierog, KIT, Germany for providing important inputs for simulating neutral pion-initiated showers and data analysis. We thank the Referee for key comments and suggestions. RKD acknowledges the financial support from North Bengal University under the Teachers' Research Project Scheme; Ref.No. 1513/R-2020.  


\end{document}